\documentclass[]{elsarticle}
\usepackage[margin=1in]{geometry}
\usepackage{hyperref}
\usepackage{epic,eepic,latexsym,amssymb,rotate,amsmath,amsthm,mathrsfs,graphicx}
\usepackage{breqn}
\usepackage{multicol}
\usepackage{natbib}
\usepackage{amsmath}
\usepackage{subfigure}
\usepackage{float}
\usepackage{caption}
\usepackage{epstopdf}
\usepackage{epic,eepic,latexsym,amssymb,rotate,amsthm}
\usepackage{multirow}
\usepackage{makecell}
\numberwithin{equation}{section}
\usepackage{hyperref}
\usepackage{booktabs}
\usepackage{parskip}  
\usepackage{multirow}
\usepackage{graphicx}
\usepackage[usenames,dvipsnames]{pstricks}
\usepackage{setspace} 

\makeatletter
\def\ps@pprintTitle{%
     \let\@oddhead\@empty
     \let\@evenhead\@empty
     \def\@oddfoot
       {\hbox to \textwidth%
        {\ifnopreprintline\relax\else
        \@myfooterfont%
         \ifx\@elsarticlemyfooteralign\@elsarticlemyfooteraligncenter%
           \hfil\@elsarticlemyfooter\hfil%
         \else%
         \ifx\@elsarticlemyfooteralign\@elsarticlemyfooteralignleft%
           \@elsarticlemyfooter\hfill{}%
         \else%
         \ifx\@elsarticlemyfooteralign\@elsarticlemyfooteralignright%
           {}\hfill\@elsarticlemyfooter%
         \else%
               {} \ifx\@journal\@empty%
                 {}%
            \else\@journal\fi\hfill\@date\fi%
         \fi%
         \fi%
         \fi%
         }%
       }%
     \let\@evenfoot\@oddfoot}
\makeatother
\begin{document}
\begin{frontmatter}
\title{Exploring Physics-Informed Neural Networks: From Fundamentals to Applications in Complex Systems}
\hypersetup{pdfauthor=Sai Ganga}
\author[]{Sai Ganga}
\author[]{Ziya Uddin}
\address{SoET, BML Munjal University, Gurugram, Haryana 122413, India}
\date{}
\begin{abstract}
Physics-informed neural networks (PINNs) have emerged as a versatile and widely applicable concept within the realm of artificial intelligence, impacting various science and engineering domains over the past decade. This article offers a comprehensive overview of the fundamentals of PINNs, tracing their evolution, modifications, and various variants. It explores the impact of different parameters on PINNs and the optimization algorithms involved. The review also delves into the theoretical advancements related to the convergence, consistency, and stability of numerical solutions using PINNs, while highlighting the current state of the art. Given their ability to address equations involving complex physics, the article discusses various applications of PINNs, with a particular focus on their utility in computational fluid dynamics problems. Additionally, it identifies current gaps in the research and outlines future directions for the continued development of PINNs. 
\end{abstract}
\end{frontmatter}
\section{Introduction}
In science and engineering, differential equations are fundamental to the mathematical modeling of several phenomena. They are crucial for explaining how systems behave when they are dependent on more than one variable and their rates of change. Almost every physical phenomena where the states are changing with respect to each other, differential equations have the capability to represent the whole  complex phenomena in terms of mathematical models, but at the same time, this complexity makes them very challenging to solve. For the simple physical phenomena the corresponding mathematical models are not too complex and the analytic solutions are possible. But in case of complex phenomena the resultant differential equations cannot be solved analytically. Therefore, numerical techniques prevails as the alternative approach to solve the differential equations for complex physics.\\
Numerous studies of solving differential equations have been conducted using traditional numerical techniques, including the finite element method, finite difference method, finite volume method, spectral methods, etc. In the literature, theoretical guarantees focusing on accuracy, stability, and convergence of the traditional numerical methods of solving differential equations are well established. These techniques have evolved over time and often satisfy the computational efficiency and robustness needed in many scenarios. However, these techniques face significant challenges as the complexity of the problem increases, primarily resulting in reduced computational efficiency. Furthermore, we might be required to rerun all of the computations if we were to make minor modifications to the initial conditions, boundary conditions, or geometry of the domain. In these techniques, the accuracy and computing efficiency of the solution are greatly influenced by the generated mesh. The mesh generation is difficult and demands a lot of computational resources. Traditional numerical discretization approaches become suboptimal when dealing with problems specified on irregular domains, problems with missing initial or boundary conditions, ill conditioned problems, or when the dimensionality of the problem increases. These drawbacks emphasize the demand for more sophisticated and flexible numerical techniques.\\
Recognizing these challenges, researchers have increasingly turned to Physics-Informed Neural Networks (PINNs) to tackle complex problems more effectively. Fundamentally, PINNs convert the challenge of approximating solutions into one of minimizing a loss function. This is done by approximating the solution using neural networks and defining the loss function such that it is the sum of the residuals of the initial and boundary conditions along with the residuals of the differential equation at certain random points within the domain. Basically, physical laws are being integrated into the neural network framework. The idea of describing loss function in this fashion without depending on any simulation or experimental data represents a significant advancement that distinguishes PINN from general neural networks.\\
Since PINNs do not require mesh generation or extensive domain expertise in numerical modeling, they provide a mesh-free alternative to established numerical methods. In addition to requiring less manual effort during simulations, PINN can effectively manage complicated simulation domains, potentially increasing computational efficiency while maintaining accuracy. In contrast to the conventional supervised learning model, PINNs is a data-free method. Nevertheless, PINNs are also capable of efficiently assimilating any available information if that is required. We should also keep in mind that PINNs may be quickly developed in modern computer languages where many packages are readily available with less effort, in contrast to the extensive stages of development of many traditional PDE solvers. A significant obstacle faced by traditional discretization methods is the curse of dimensionality, which was initially described by Bellman and Kalaba \cite{rev84}. High-dimensional differential equations are utilized in several academic disciplines, including engineering and finance. In these kinds of situations, PINN is very efficient. Apart from the aforementioned benefits of PINNs, the solution obtained using neural network approximation has various beneficial attributes. These include their exceptional interpolation capabilities, minimal parameter count, appropriateness for efficient parallel computer implementation, and smooth and differentiable nature, which makes them suitable for use in optimization problems involving differential equation residuals.\\
Despite its innovative features, PINNs has certain inherent restrictions. Given its recent development as a numerical approach, PINN has a less rigorous theoretical foundation and a dearth of effective error analysis tools. Understanding the convergence criteria in the optimization process, where solutions can get trapped in local minima, can be challenging in PINNs since the given loss function can often be non-convex in nature. Also, integrating several objectives into the loss function during the training of the network can lead to the gradients being biased.  PINNs are similarly prone to vanishing gradient problems, especially when the networks are deep. Neural network training often demands longer periods of time, depending on the selected architecture, solvers under consideration, etc. It is not simple and straight-forward to train a neural network since PINN does not always converge, and optimizing hyper-parameters requires a lot of time and effort. Also, which framework has to be chosen cannot be ascertained for a particular problem.\\
Nonetheless, recent developments have tackled the significant challenges of PINNs, strengthening their relevance in many areas of science and technology. As the study of PINN evolves, it has the potential to be a vital tool for real-world uses. PINNs role is definitely not to replace the classical numerical solvers, while PINNs are well suited for simulating high-complexity industrial problems where conventional approaches fail.\\
As previously discussed, PINNs are highly effective in addressing differential equations that describe complex physical phenomena. With the ongoing technological advancements driving the world towards energy-efficient systems and sustainable energy solutions, effective thermal management in machinery and industrial processes has become a crucial, albeit indirect, objective. Thermal management in both industrial and natural settings involves intricate fluid dynamics, as well as heat and mass transfer, making accurate simulation of fluid flows essential. Computational fluid dynamics (CFD) presents numerous challenges, necessitating the explicit characterization of fundamental physical principles to accurately study fluid flows and heat and mass transfer. Furthermore, fluid flows often exhibit complex, multiscale behaviors, requiring algorithms capable of managing these complexities. In recent years, several innovative PINN algorithms have been developed, offering significant potential for applications in fluid dynamics and dynamics and beyond.\\
To contextualize the ongoing research efforts, we document a few review papers available in the literature. Karniadakis et al. \cite{rev39} reviewed the PINN methodology, its diverse applications, and its current strengths and limitations. A review of PINN for fluid dynamics is done by Cai et al. \cite{rev40}, where recent advancements of PINN were reviewed briefly. A few case studies were also carried out to show the potency of PINN in which inverse problems were solved using PINN with the availability of limited data and physics information. The comprehensive analysis of the innovation process in the area of PINN from 2018 to 2021 can be found in \cite{rev43}. Faroughi et al. \cite{rev50} reviewed the methodology, application, and limitations of four different kinds of neural networks, including PINN, physics-guided neural networks, physics-encoded neural networks, and neural operators. Antonion et al. discuss the advancements and difficulties with PINN and its variants \cite{rev51}. A bibliometric analysis of PINN can be found in \cite{rev52}. The review of the advancement of PINN in terms of strategies, representation of the available physics information, and methods to incorporate it is provided by Hao et al. \cite{rev82}. Different PINN techniques, the mathematical aspect of the software package NeuralPDE.jl, and its applications to different PDEs are extensively discussed by Zubov et al. \cite{rev53}.\\
This article explores the evolution of PINN as a highly promising numerical method. We discuss the method's initial successes, identify its limitations, identify improvements made to address these limitations, highlight the ongoing challenges that remain to be addressed, and discuss the applications of PINN to thermal management systems and fluid flow. This comprehensive review of the evolution of PINN is expected to provide a valuable resource for researchers and practitioners in the field to apply PINN across diverse disciplines. The main theme of this review paper is to understand the importance and need for the development of PINN.
\section{Fundamentals of PINN}
A neural network is an artificial intelligence technique that trains machines to process information in a manner similar to that of the human brain. Neural networks offer significant benefits when it comes to the approximation of a function and act as universal approximators \cite{rev85,rev86}. The work by Hornik et al. \cite{rev86} “rigorously establishes that standard multilayer feedforward networks with as few as one hidden layer using arbitrary squashing functions are capable of approximating any Borel measurable function from one finite dimensional space to another to any desired degree of accuracy, provided sufficiently many hidden units are available.” There are different types of neural networks, including feed-forward neural networks, convolutional neural networks, recurrent neural networks, generative adversarial networks, etc. Although feed-forward neural networks are the focus of most investigations, other PINN extensions based on different types of network architectures \cite{rev43} have also been studied. A neural network can receive single or multiple inputs. It is also capable of producing single or multiple outputs. A neural network is composed of weighted linear summation and activation functions in between the input and output. These processes take place on a neuron. So, in a neural network, there is an input layer, multiple hidden layers, and an output layer. In the PINN settings, the solution of the differential equation is approximated using the neural network. If we assume that $z^i$ represents the output of a node of the $i^{th}$ hidden layer, then the neural network can be written in the following way:
\begin{flalign}
& \begin{aligned}
& z^0 = x = (x_0,x_1,\dots)   \\
& z^i =g(W^{i}z^{i-1} + b^i) \quad \text{for} \quad 1 \leq i \leq N-1 \\
& z^i = W^{i}z^{i-1} + b^i \quad \text{for} \quad i=N
\end{aligned} &&
\end{flalign}
Let us assume that the approximated solution of the considered differential equation is $\widehat{f}$. It is obtained from the final output of the neural network $\widehat{f} = z^N$. The input with multiple coordinates is given by $x = (x_0, x_1,\dots)$. If we are considering a PDE with space and time coordinates, then it would be $(t, x)$. The values of the input coordinates are chosen from the domain of the definition of the problem. In most studies, inputs are randomly initialized. $W^i$ represents the weight matrix of the $i^{\text{th}}$ layer. Inclusion of weights in the neural network approximation could improve the approximation and accuracy. $b^i$ represents the bias vector of the $i^{\text{th}}$ layer. In neural networks, the inclusion of a bias vector might improve the network's generalisation and adaptability. The trainable parameters of the neural network are the weight matrix and the bias vector. These parameters are generally initialized randomly using a normal or uniform distribution or utilizing Xavier/Glorot \cite{rev87} uniform or normal initialization techniques. $g$ represents the activation function. The purpose of an activation function is mainly for a nonlinear transformation of the weighted sum. A nonlinear activation function is essential for making the neural network more complex. The accuracy and convergence of a PINN greatly depend on the activation function selected, and utilizing different activation functions offers notably varying performance. A few of the activation functions that are generally used are sigmoid, tanh, softplus, and swish. In the upcoming sections, the latest advancements made to address some of the shortcomings of the available activation function choices and initialization methods will be discussed. The crux of PINN lies in the minimization of the loss function. A key factor in determining what the network learns to optimize during training is the loss function. The PINN loss function is defined as the sum of the residuals of the differential equations and their boundary or initial conditions. The most commonly used norm to define the residual is $L^2$ norm; however, there have been studies on its influence, which will be discussed in the coming sections. Another important piece of research was on the impact of weights that have to be multiplied to the individual terms of the loss function, which is also discussed in the coming sections. To show the how the loss function is defined, let us consider a general partial differential equation as:
\begin{flalign}
& \begin{aligned}
& \mathcal{L}[f(x)] = \phi(x)  \quad \forall \quad x \in \Omega \\
& \mathcal{B}[f(x)] = \psi(x) \quad \forall \quad x \in \partial\Omega \\
\end{aligned} &&
\end{flalign}
Here, $x$ is considered as the independent variable, $f$ is the solution, $\phi$ and $\psi$ are functions of x, $\Omega$ is the domain, $\partial\Omega$ the boundary, $\mathcal{L}$ represents the partial differential operator, and $\mathcal{B}$ represents the boundary condition.
Let $\Omega_f$ represents chosen $N_f$ collocation points from $\Omega$, and $\Omega_b$ be $N_b$ points from $\partial\Omega$. The value of $N_f$ and $N_b$ is user defined depending on the problem. Now we define the PINN loss $L$ as the summation of the $L^2$ norm of the residual of considered PDE and its boundary conditions resulting in definition of the loss function in mean squared error sense:
\begin{flalign}
& \begin{aligned}
& L = L_f + L_b \\
& \text{where,}\\
& L_f = \frac{1}{N_f}  \Big\| \mathcal{L}[f(x)] - \phi(x) \Big\|^2 \quad \text{for} \quad x \in \Omega_f\\
& L_b = \frac{1}{N_b}  \Big\| \mathcal{B}[f(x)] - \psi(x) \Big\|^2 \quad \text{for} \quad x \in \Omega_b
\end{aligned} &&
\end{flalign}
Fundamentally, in order to train the neural networks, the loss function must be minimized. Back propagation offers an efficient way to calculate gradients and carry out gradient-based optimization. The process of neural network training involves using the gradients calculated by back propagation to iteratively modify the network's parameters. Many optimizers have been used by the researchers, which play a major role in the training process. A few of them include stochastic gradient descent, limited-memory BFGS \cite{rev89}, and Adam \cite{rev88}. Choosing the right optimizer is a critical component of PINN for not getting stuck at the local minima and to enable higher accuracy and convergence. It is necessary to calculate the derivatives of the network outputs with respect to the network inputs and the derivative of the loss with respect to the trainable parameters when carrying out the gradient descent. We can utilize options like manual calculation of derivatives, symbolic differentiation, or numerical differentiation. However, solving the derivatives by hand isn't very practical; symbolic differentiation will cost a lot of memory and can be slow because of the resulting massive expressions; and in numerical differentiation, round-off errors can occur. In such scenarios, automatic differentiation \cite{rev90} is an alternative to these methods for computing the derivatives. It employs precise expressions with floating-point values in place of symbolic strings without any approximation error. Automatic differentiation computes the derivatives by applying the chain rule. In automatic differentiation, there is a forward pass to calculate all variable values and a backward pass to calculate derivatives. Automatic differentiation libraries are available in a number of computational frameworks, making the process easier. Automatic differentiation has greatly advanced PINNs from a promising idea to an effective tool for solving complex problems. However, inefficiency and instability that might occur while using automatic differentiation to evaluate the $n^{\text{th}}$ order derivatives can lead to inefficiency and instability \cite{rev31}, where an exploration of new advancements might be necessary. The schematic representation of genral PINN for the considered PDE is provided in Figure 1.
\begin{figure}[H]
    \centering
    \includegraphics[width=16cm, height=8cm]{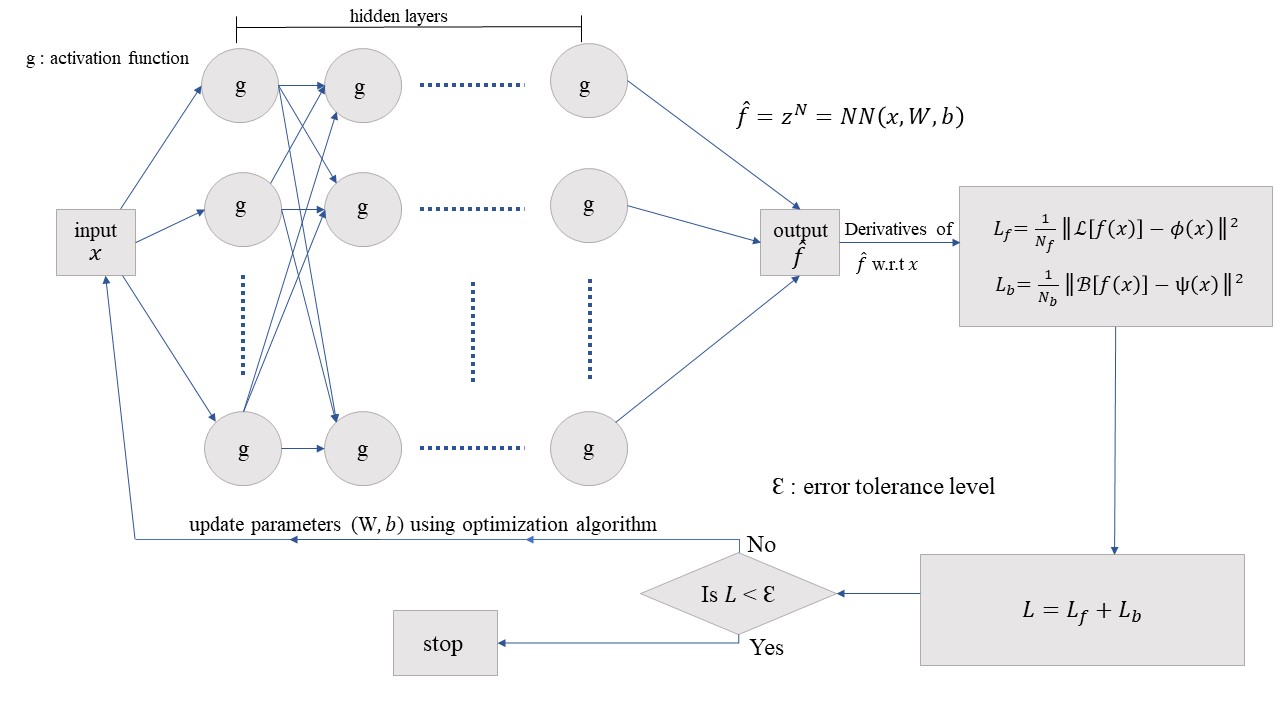} 
    \caption{Schematic representation of PINN}
\end{figure}
\section{Pathway to PINN}
The idea of solving the differential equations (DEs) by transforming it to an unconstrained minimization problem is not new. The reason why the study of PINNs has seen a significant upsurge in recent years is mostly due to the modern advancement in methods and algorithms allowing it to solve a wide class of problems. One of the earliest works in demonstrating the use of neural network in approximating the solution of partial differential equations (PDEs) is given by Dissanayake and Phan-Thien \cite{rev0}. In their work they demonstrate the simplicity of using neural network's approximation capabilities to find the solution of non-linear and inhomogeneous PDEs. 

Amongst the earliest work, another significant work was done by Lagaris et al. \cite{rev5}. In their work the solution of differential equation is approximated by a trial function which is written as the sum of two terms, in which one term satisfy the boundary condition and the other term is approximated using the neural network. Since the constraints are satisfied by this approximation, we are left with a simpler unconstrained optimization. However, the defined trial function will not be capable of approximating DEs defined on irregular boundaries as the method is defined for DEs with orthogonal box boundaries.

Later this methodology was improved by Lagaris et al. \cite{rev8} which could address the boundaries with complex geometry. A multilayer perceptron (MLP) that approximates the solution over the domain including the boundary is first created using the penalty function formulation. Next, to the obtained MLP network, another network named a radial basis function (RBF) is applied. The precise satisfaction of the boundary conditions is taken into consideration while using the RBF approach.  The technique that was presented could handle boundary value problems with irregular boundaries, however it only emphasised on problems with Dirichlet and Neumann conditions.

The term physics-informed neural networks (PINNs) were first introduced in 2017 in the works of Raissi et al. \cite{rev6, rev6.1}. Forward problems and inverse problems of non-linear partial differential equations were addressed in the first and second parts, respectively. A combined version of these two articles was subsequently published in 2019 \cite{rev6.2}. The authors presented two classes of algorithms, including continuous and discrete time models, and used several benchmark problems to illustrate the characteristics and effectiveness of each class. It was noted that the first class presented might face difficulties in higher-dimensional problems since it would need a lot of collocation points to impose physics on the loss function. To overcome this problem, an alternative method using the traditional Runge-Kutta time-stepping approaches was employed, which would give a structured neural network representation.

PINNs have a number of shortcomings, which has led researchers to investigate various adjustments in the methodology in order to improve its effectiveness and broaden its potential for addressing real-world problems. Now we will discuss the modification of PINNs and different methodological developments of PINNs. 
\section{Modification of PINN inspired from traditional techniques}
Kharazmi et al. \cite{rev20} introduced a variational physics-informed neural network (VPINN), which establishes a new loss function in variational form. The variational form is obtained when the PDE residual is integrated over the defined domain after multiplying it with the test function. This variational residual gives rise to variation loss. The solution is approximated using neural networks. This method combines PINN with the traditional Petrov-Galerkin method. The function space of neural networks is taken as the trial space, and the function space of Legendre polynomials is taken as the test space. VPINN with the delta dirac function as the test function is essentially PINN. VPINN provides several advantages over PINN. Training cost and accuracy are better in VPINN, as the differential operator’s order is reduced when we integrate the variational form. However, the computation of the analytic form of this integral can be difficult if we consider a more complex neural network structure. In specific cases, a shallow network was utilized and the analytic expression was obtained, which is advantageous for carrying out numerical analysis of the method. But for the cases with deep networks, the numerical integration technique was utilized. The difficulty here is that there is no specific quadrature rule designed for the numerically integrating the function involving deep neural networks. However, the number of quadrature points that we need to be using is smaller as compared to the large number of collocation points that we might need to take for PINN. 

Khodayi-mehr and Zavlanos \cite{rev21} formulated the VARNET library for solving PDEs using VPINN. Additionally, an optimal way of choosing the training points rather than choosing them arbitrarily is proposed in the work, which is also incorporated in the VARNET library. Adding a larger number of training points where the solution varies will enable neural networks to learn them accurately. For the same, a residual field of the considered PDE is utilized. PDE residual field is basically substituting the neural network approximation for a specified set of input data into the PDE. Given the residual field, a rejection sampling algorithm is employed to sample the spacetime. The specified sampling algorithm will determine the sampling points that give a higher residual value. The concept of local adaptive mesh refinement is comparable to this method of choosing optimal training points. This strategy makes the training process more efficient and gives better prediction accuracy.

hp-VPINN is an extension of VPINN, where the domain decomposition is also taken into consideration. The methodology is the same as in VPINN; the difference lies in the fact that the domain is decomposed into several sub-domains where the test function is locally defined over each sub-domain. This method combines PINN with the sub-domain Petrov-Galerkin method. The test space consists of the piecewise polynomials that do not overlap on each sub-domain. As the approach is based on domain decomposition, if we focus on the sub-domain having a less regular solution, we will be able to improve the training cost and achieve a more localized learning process. The advantages of hp-VPINNs in solving differential equations are demonstrated in \cite{rev22} for several examples. The method's adaptability enables it to handle steep solutions, singularities in solutions, and sharp changes in solutions.

Another improvement of PINN is conservative physics-informed neural network (cPINN). This methodology can be applied for nonlinear conservation laws. In this formulation domain decomposition is utilized where separate PINN is utilized in each sub domain. All these sub domains are then joined together using interface conditions. The flux continuity is considered at the interfaces of each sub-domain. In addition to this, an average solution is also considered at the common interface between two sub-domains, it is observed that adding this additional interface condition gave a better rate of convergence. An advantage of having domain decomposition is that, we can define complex neural network structure or simple neural network structure depending on whether or not we have smooth solution in individual sub domains. Another advantage is that domain decomposition facilitates parallelized computing effectively. Additionally, by allocating a distinct network to each sub-domain, cPINN can deal with piecewise constant coefficients, making it more adept at handling inverse problems than PINN. There are various other advantages which are discussed in detail in \cite{rev17}, in which various test cases are also solved. 

cPINN can be extended to solve any kind of DEs such that the interface conditions depend on the governing equation of the problems that we are considering rather than those coming alone from the conservation laws. Extended physics-informed neural networks (XPINNs) \cite{rev18} is one such method. The primary benefit of utilizing domain-decomposition based approaches is their ability to tackle one of the major challenges of PINN while we apply it to real world problems, which is higher training time. XPINN comes with all the advantages of cPINNs with additional advantages of generalized decomposition of space-time domain and simpler interface conditions.

Hu et al. \cite{rev19} provides theoretical and numerical explanations of how well XPINN performs in comparison to PINN, with particular emphasis on when and how to choose XPINN over PINN and vice versa. In particular, the work elaborately presents the trade-off in the generalization of XPINN. XPINN performs better since it is having a complex network only on some sub domains and not over the entire domain. However, PINN outperform XPINN when there is an over fitting because of the less training data. So, there are two aspects of the trade off, the simplicity because of the domain decomposition and the lack of training data which can cause over fitting. XPINN performs better than PINN when the former is more prevalent. If not, PINN performs better than XPINN. XPINN and PINN perform equally once the two components behave in a balanced way.

Meng et al. \cite{rev23} introduced parareal physics-informed neural network (PPINN). For implementing the approach, the PDE is simplified as a special case of the exact PDE, and the coarse-grained solver (CGS) solves this simplified PDE sequentially across the entire time range to get an initial solution. Then, the domain is divided into $N$ sub domains where the exact PDE is solved using $N$ fine PINNs. The initial conditions required for the fine PINNs in all of the sub domains are obtained from the solution provided by the CGS.  PPINNs is a predictor corrector approach, where the predictor is the simplified PDE that the CGS solves and the corrector is the exact PDE that fine PINNs solves. The advantage of this method is that since we are using fine PINNs that are trained using smaller data sets, training time is reduced, also parallelization of this fine PINNs can be accomplished. Also, if the CGS performes efficiently, within a few iterations this method will converge. This parareal algorithm (PPINN) are better than the existing parareal algorithms. This is because, using PINN we can easily solve problems with even partial information but existing methods will find it difficult to deal with PDEs with partial information.

A hybrid physics-informed neural network (hybrid PINN) was devised by Fang \cite{rev29} inspired by the finite volume method and convolutional neural networks. One of the hurdles in the PINN study is its lack of theoretical understanding. In a few of the numerical examples, it was observed that even if the loss function reaches zero, PINN often gives wrong predictions. Trying to fix this issue is the main motivation behind proposing the hybrid PINN methodology. For the same in an arbitrary geometry, an approximation of the differential operator was proposed, which is termed the local fitting method. Basis functions like polynomial functions that satisfy certain conditions are chosen to approximate the differential operator, which is written as a linear combination of these basis functions. Then, the coefficient of the linear combination is determined using the least squares method. The solution of the PDE is approximated using a neural network. So, to evaluate the residual of the PDE, we can make use of the approximation of the differential operator, which is almost like the convolution operation that arises in convolutional neural networks. That is, the employment of automatic differentiation is replaced with this approximation. The main advantage is that the proposed algorithm promises a convergent rate, which has been proved theoretically. Another benefit mentioned is that it comes with the flexibility of PINN but with better accuracy. Also, for approximating the differential operator when considering a fixed geometry, the same matrix can be used, which makes it computationally efficient and apt for transfer learning. The disadvantages of the method in some scenarios and the possible improvements are also discussed in detail, including the stability of the method and its difficulty when it comes to non-linear operators.

Markidis \cite{rev47} combined the strengths of PINN and traditional linear solvers to obtain the solution of the Poisson equation. This integration is a practical option to consider because the convergence of PINN to the low-frequency components of the solution is faster, whereas for the high-frequency components it is slower. Because traditional solvers may converge faster to the high-frequency components of the solution, we can thus integrate PINN with them, thereby combining the strengths of both methods. Initially, on a coarse grid, the solution of PINN is approximated. And then the Gauss-Siedel method is utilized for refining the obtained solution on a coarse grid. Finally, it is linearly interpolated to finer grids, to which Gauss-Siedel iteration is applied. This work also emphasizes the vital significance of applying transfer learning.
\section{Methodological developments of PINN}
\subsection{Loss function}
Yu et al. \cite{rev24} formulated gradient-enhanced physics-informed neural networks (gPINNs). This methodology is seen to perform better than PINN in terms of training time and accuracy. The idea of gPINNs is that since the residual of the DEs is zero, the derivative of the residual will also be zero. So, in the loss function the gradient of the residual is added in addition to the terms that are generally present in PINN loss function. Various test cases were solved to show the effectiveness of the method. It was shown that performance  of gPINN significantly enhanced when residual-based adaptive refinement was incorporated. The residual-based adaptive refinement approach adds more residual points to the network wherever a high PDE residual occurs during training. Due to the additional loss terms, gPINN has a larger computational cost than PINN. This is one of its drawbacks. Additionally, the weight present in the loss term containing the gradient information is seen to influence the performance in some problems, therefore it would be necessary to automatically choose an ideal weight.
\subsection{Activation function}
The importance of the activation function in PINN is extensively discussed in the literature. One of the important works about an improved activation function is by Jagtap et al. \cite{rev28} where theoretical proof is provided, which roughly states that using an adaptive activation function with a suitable initialization process and a proper choice of learning rate doesn’t let gradient descent methods converge to local minima. Apart from improved convergence, the adaptive activation function shows improved accuracy as well. This is shown with the help of various examples. In the adaptive activation function, an additional hyper parameter along with a scaling factor is introduced in the activation function, which can be optimised dynamically during the optimization process. This kind of hyper parameter has the ability to alter the activation function’s slope, which is cruicial in the training of neural networks.

Uddin et al. \cite{rev44} studied the impact of various activation functions in approximating various non-linear differential equations, including PDEs, coupled equations, and DEs defined over an unbounded domain. In their work, different wavelet activation functions were utilized to solve the problems, and they were compared against the tanh activation function. It is observed that wavelet activation gave better accuracy for all the considered problems when compared with tanh activation. However, it was concluded that which wavelet activation function performs best depends on the problem under consideration. In their work, different neural network approximations were utilized for approximating different solutions when dealing with coupled equations, which was seen give better performance. An analysis of network architecture is also carried out for the considered problems.

Utilizing non-Newtonian fluids Yang et al. \cite{rev48} constructed a traffic flow equation and developed a methodology using a physical-informed rational neural network. They also developed an optimization algorithm named noise heavy-ball acceleration gradient descent. In the PIRNN framework, a rational activation function is utilized owing to its improved approximation property, which is seen to enhance the performance of training. A rational activation function is a nonlinear activation function that is a rational function having trainable polynomial coefficients on both the numerator and denominator. A brief note about the new optimization strategy is given in the next subsection. In the paper, it was concluded that this methodology of PIRNN, along with the new optimization strategy, was performing better than traditional PINN and the finite element method.
\subsection{Optimization algorithms}
Sabir et al. \cite{rev1} solved the Lane-Emden equation, which has a singularity at the origin, by approximating the solution using neural networks and utilizing the strength of the meta-heuristic algorithm. In this work, the optimization algorithm utilized is a combination of the genetic algorithm and the interior point algorithm, making use of the Morlet wavelet as the activation function. The same methodology was utilized by Sabir et al. \cite{rev2} for higher-order non-linear Lane-Emden equations having singularities at the origin. Another similar work in this direction in solving fractional non-linear Lane-Emden equations having singularities at the origin can be found in \cite{rev10}. In this work, a genetic algorithm is combined with an active set algorithm using a Meyer wavelet kernel. Ilyas et al. \cite{rev4} applied a combination of genetic algorithms and sequential quadratic programming as the optimization algorithms and approximated the solution of the Falkner-Skan equation using a neural network in which Gaussian wavelet kernels were utilized as activation functions. In all of the above works statistical analysis is carried out to validate the convergence and accuracy. It should be mentioned that these methods are effective and easier to implement. Additionally, solving Lane-Emden models with conventional approaches might be difficult; in these situations, applying meta-heuristic-based algorithms and employing neural networks to approximate the solution is shown to be an appropriate strategy. 

Tan et al. \cite{rev9} solved elliptic PDEs by using a modified metaheuristic algorithm for optimization, namely an improved butterfly optimization algorithm. The solution of the PDE is approximated using a trial function, in which one term is approximated using neural networks. The trial function is defined in such a way that it satisfies the boundary conditions. They also utilized a Gaussian wavelet as an activation function. While comparing their results with other metaheuristic algorithms used for optimization, they concluded that the methodology proposed in this work gave better accuracy. 

As mentioned earlier, utilizing non-Newtonian fluids, Yang et al. \cite{rev48} constructed a traffic flow equation. In their newly developed methodology, they also introduced an optimization algorithm named noise heavy-ball acceleration gradient descent. This method is a non-convex optimization algorithm that is observed to be very efficient. The iterative format of the gradient descent is described in the work. This method merges the strengths of the noise gradient descent method and the heavy ball method, which are the global optimization process and the fast acceleration of the optimization process, respectively.
\subsection{Network architecture and hyper parameters}
Wang et al. \cite{rev37} address one of the major challenges of PINN, which is the imbalance of the gradients that occurs while training the model. The reason for this imbalance resulting in the failure of PINN is analyzed and identified in detail. The main reason for the failure of PINN is due to the fact that the training of the model has a higher bias towards the residual term. This happens when the gradient of the boundary conditions is smaller during training and that of the residual term is larger. In the former case, boundary data cannot be appropriately fitted, and in the latter case, the model will learn those solutions that fit the DEs correctly. This imbalance of the terms in the loss function leads to an incorrect prediction of PINN. This challenge is addressed using the learning rate annealing approach, which balances the terms in the loss function during the training process. Additionally, a new neural network architecture is also presented, which is less stiff than the traditional fully connected neural network. The gradient flow stiffness is also analyzed and quantified in the work. The validation of the ideas is carried out with the help of various problems, and an improvement in accuracy is established. The proposed learning rate annealing technique is inspired by the ADAM algorithm, which is intended to adjust the weights dynamically using the gradient information obtained while the model is being trained, such that all the terms in the loss function are balanced. The new architecture draws inspiration from the neural attention process, which possesses specific features including residual connections and consideration of the multiplicative interactions between inputs.

One of the biggest challenges of PINN is finding the optimal hyperparameters. The efficacy and precision of PINN are also highly influenced by the hyperparameters. Wang et al. \cite{rev49} introduce Auto-PINN, which automates the process of determining the hyperparameters rather than the tedious process of manually figuring out the right set of parameters. This technique makes use of a neural architecture search. The idea utilized for the search is that, initially keeping one hyperparameter varying and all others fixed, the ideal hyperparameter that is varying is determined, and then the obtained choice is fixed and other parameters are searched. This helps to reduce the scale of the search space. So, a decoupling approach is used in a step-by-step manner. For the methodology, the search objective is considered to be the loss function, as from the experimental observations it is concluded that there is a correlation between the errors and the loss function. They also did a comparison of Auto-PINN with baseline methods, including Random Search and HyperOpt, on various PDEs and obtained a better performance result. Furthermore, an analysis of the effects of epochs, data sampling methodologies, and learning implied that optimal configuration was influenced by the sampling method, but epochs and learning rates didn’t have much impact. Another effort in this direction was given by Escapil-Inchausp{\'e} and Ruz \cite{rev60}, where Gaussian processes-based Bayesian optimization was employed for hyperparameter optimization (HPO). They applied HPO to the Helmholtz equation to demonstrate the application and need of the method. HPO offers an effective approach to address the issues of PINN's slow training and its difficulty in finding appropriate configurations due to the large dimensionality of the hyperparameter search space. Wang and Zhong \cite{rev61} propose the neural architecture search-guided physics-informed neural network NAS-PINN. In essence, NAS-PINN is the integration of NAS into the PINN framework. NAS allows for the identification of the best neural network design within a specified search space. A bi-level optimization is used to train NAS-PINN, in which unknown parameters to be determined, including weights and biases, are optimized by the inner loop and architectural hyperparameters are optimized by the outer loop. By performing a number of numerical tests on some of the benchmark problems, the capability of NAS-PINN is demonstrated. Zhang and Yang \cite{rev62} make use of an evolutionary algorithm to determine the best PINN model. The method simultaneously examines the structure of the neural network and the activation function. They have also adopted dynamic population size and training epoch techniques in evolution. This utilization is said to greatly expand the pool of models to search for and speed up the process of finding models with a high rate of convergence. By solving several numerical examples and comparing the model obtained using the proposed methods to other existing methods, including random search, Bayesian optimisation, and evolution without using the adopted technique, it was concluded that the proposed model gave better precision and rate of convergence. The limitations and future improvements are also mentioned in the work. A few of which are as follows: Firstly, defining the loss function as the objective of evaluation does not necessarily ensure that it appropriately reflects the error of the model. Secondly, how the activation function performs is dependent on how the neural network’s initial parameters are distributed, but initialization of the parameters specific to different activation functions is not carried out.

Gladstone et al. \cite{rev54} formulate FO-PINN. In the proposed method, if we have to solve a $d$-order PDE, all of the $d-1$ derivatives of the solution are the neural network output, and a set of compatibility equations is also included in the loss function, which is basically the error of exact and predicted derivatives. This improves the training speed and accuracy since the computation of higher-order derivatives is not required. Using the approximate distance function, the proposed method is capable of imposing exact boundary conditions, which further improves the accuracy, which isn’t trivial using traditional PINN. Additionally, since PINN involves higher-order derivatives, using Automatic Mixed Precision (a method of training) is not applicable, whereas its applicability is possible in FO-PINN, which further improves the training process. These advantages of FO-PINN are demonstrated with the help of numeric examples.

A self-training physics-informed neural network (ST-PINN) is introduced by Yan et al. \cite{rev55}, aiming to address the issue of convergence and reduced accuracy of PINN. The loss is evaluated after a few iterations, then a sample of points that give the least loss value is chosen, and pseudo-labels are given to those points. This obtained information is fed into the loss function as available supervised information. This is repeated for the following iterations. This strategy has been observed to provide better convergence. The advantages of the proposed method are exemplified by solving five PDEs. There are certain problems with the current methodology. Firstly, in this method, choosing a threshold is often challenging as it is problem-dependent. Secondly, since the pseudo points are generated in every iteration, it affects the method's efficiency, so the frequency at which the pseudo points are updated can be reduced. Thirdly, the addition of pseudo points can affect the training process and will lead to fluctuations, leading to instability of the training process, so it is required to check the count of pseudo points to be included. However, these three problems are addressed in the paper, respectively, by introducing three hyperparameters, namely the maximum rate, the update frequency, and the stable coefficient. 

Hu et al. \cite{rev65} introduced a new method having a lot of advantages over the existing PINNs to solve high-dimensional PDEs, namely stochastic dimension gradient descent (SDGD). In standard PINN training for high-dimensional PDEs, the loss function will contain large PDE terms with different dimensions. Computing the gradient of this entire term is often time- and memory-consuming. In the proposed method, the gradient of the residual is decomposed into different components, where each component represents a different dimension of PDEs. So, the proposed method will accelerate PINN training. After the decomposition, for every iteration, a subset of the decomposed components is sampled for the optimization process of PINN. This sampling is not observed to be biased and assures convergence. The convergence and other characteristics of the entire algorithm are proved theoretically, and the performance is demonstrated by applying it to various problems. A detailed comparison of the proposed method to standard PINN is also carried out in the study. The proposed method has several advantages, including a low cost of memory, good accuracy, faster convergence, gradient computation is unbiased of the full batch gradient, the method can be generalized to any arbitrary PDE, and the solution can be predicted over the entire domain. The method also comes with certain limitations. Firstly, if batch sizes that are too small are used, it can slow down the convergence; hence, for very high-dimensional PDEs, batch sizes that are often larger in size have to be utilized. Secondly, if the loss function has to be defined differently, for example, in Schr\"{o}dinger’s equation, even though the proposed method is flexible enough to solve the problem that cannot be addressed by standard PINN, it can cause problems regarding the memory cost.
\section{Theoretical advances on PINN}
One of the major limitations of PINN is its lack of strong theoretical foundations. However, the success of PINNs in their applicability to a wide range of problems has attracted many researchers to perform theoretical analysis. Some of the theoretical works of PINN are discussed below. Shin et al. \cite{rev67} considered linear second-order parabolic and linear second-order elliptic PDEs to show the consistency of PINN theoretically. An upper bound of the PINN loss is computed by using probabilistic space-filling arguments. The uniform convergence of the sequence of minimizer to the solution of the PDE was also proved by using the Schauder approach. This may alternatively be understood as the convergence of the PINN's generalization error to zero. According to the analysis of the considered problems, the convergence of PINN is assured with the satisfaction of certain hypothesis, and the theory of convergence is given in terms of the number of training data. The theoretical results were also demonstrated with the help of examples. 

Upper bounds of PINN’s generalization error while approximating the solutions for forward problems are provided by Mishra and Molinaro \cite{rev68}. That is, an abstract framework for deriving error estimates for PINNs is studied rigorously. The work is similar to the previous study provided in \cite{rev67}. However, the present approach is not restricted to specific PDEs; any general nonlinear PDEs are covered in the study. Also, the generalization error is estimated directly using the described stability estimate and the loss function. The study assures a smaller generalization error if we have a small training error with sufficient quadrature points having accuracy on the provided quadrature rule, and there is some oversight over the constants pertaining to the given PDE stability. Several examples are also provided to demonstrate the proposed theory. However, this study is subject to certain limitations, including that the provided methods won’t be able to address PDEs with a rough solution, and the derived estimate depends on training error, which cannot be rigorously estimated. But if such estimates can be computed, then they can be included in the derived estimate. Additionally, the present study deals with the estimates only for forward problems for PDEs. \cite{rev69} contains an extended version of this work that provides thorough estimates for PINN’s generalization error for inverse problems. In these studies, if the sampling points and the neural network parameters go to infinity, convergence is not assured.

Shin et al. \cite{rev71} conducted a similar study of determining the estimates of error that minimize the loss function for generic linear problems covering elliptic, parabolic, and hyperbolic PDEs. Additionally, different types of loss functions are considered depending on whether the residual is defined in strong or weak form. The definition of the weak form of residual is discussed priorly \cite{rev20,rev21,rev22}. So, the current study can be applied to the recent methodologies for PINN extensions. For these two types of loss functions, error estimates are derived for discrete as well as continuous cases, and under certain hypotheses in the defined topology, convergence is established. In addition, these theoretical findings are demonstrated with the help of various test cases. There are certain limitations in this work that are mentioned in the study, and thus further development is essential. One is that when it comes to multiscale problems, there are many factors that have to be considered, for example, the inclusion of essential physical properties. Secondly, the method is considered only for linear problems and not for nonlinear problems. Though nonlinear problems were considered in \cite{rev67}, the study can still be extended in such a way that defined hypotheses are more flexible. An overview of related research is provided by \cite{rev71}.

Wang et al. \cite{rev70} explored whether defining the loss function using $L^2$ norm is a suitable choice for the training process to give a better approximation for the solution of the PDE. It should be noted that if, with respect to the defined PINN loss function, the stability of PDE is not guaranteed, then a precise solution cannot be obtained. Theoretical study is carried out inspired by the idea of proving the stability of a numerical scheme. So, in the PINN framework, the stability theory is analyzed. That is, in the case of PINN, as the loss tends to zero, the asymptotic behavior of the difference between the exact solution and the predicted PINN is analyzed. The theory is carried out for the Hamilton Jacobi Bellman (HJB) equation, which is a high-dimensional PDE that is nonlinear in nature. For this class of PDEs, it is proven that if we are considering the standard $L^p$ loss, the stability of this equation is attained only for large values of $p$. This result throws light on the fact that for small values of $p$, that is, if we consider $L^2$ loss for the HJB equation, then even if we are getting a small loss value, the predicted value of the solution need not be close to the exact solution, and therefore it would be ideal to define $L^{\infty}$ loss. The fact that $L^{\infty}$ loss acts as a good loss can be confirmed by the theorem stated and proved in the study. Numerical experiments are also carried out to demonstrate the claim. As the majority of the PINN study is based on $L^2$ loss, this study also introduces a training strategy for HJB equations such that $L^{\infty}$ loss is minimized. The new training strategy adopts a min-max optimization process. The min-max algorithm works on two different loops: the outer loop, which is a minimization problem, and the inner loop, which is a maximization problem. The inner loop determines those points where the predicted solution substantially differs from satisfying the PDE. The outer loop finds the best parameters such that the loss is minimized. The gradient of the model parameters is evaluated after inner-loop optimization is over. Any optimization algorithm can be used to update the parameters. The advantage of the theoretical development and new training strategy conducted in the study is that after the training procedure is completed, we can assure that the predicted solution will be very close to the exact solution if the loss obtained is small.

Even though the application of PINN has been very successful, not much exploration has been carried out to understand how PINN training is successful and why it fails at times. An important study comprehending PINNs and the dynamics of their training was carried out by Wang et al. \cite{rev72}. With the help of the neural tangent kernel, they investigate this and provide a theoretical understanding. Empirical experiments were also performed to validate and demonstrate the theory.

Lu et al. \cite{rev31} developed DeepXDE, which is a Python library used for implementing PINN. Implementation of this library is easy, and the demonstration of its usage and its customizability is also explained in their work, in addition to demonstrating a few examples showing the capability of the library and the efficacy of PINNs. More importantly, the theory of approximation and the analysis of error are also explained. To further improve the training, a residual-based adaptive refinement technique is also introduced, which is a new way to choose residual points while training rather than selecting them randomly. This method is observed to be efficient in most cases, except for those PDEs whose solution has steep gradients. 
\section{Applicability of PINNs in solving DEs}
PINN can be widely applied to solve various kinds of differential equations. Pang et al. \cite{rev77} introduced fractional PINN (fPINN). This methodology is capable of handling operators that are of fractional order and operators that are of integral order. It should be stressed that PINN cannot handle fractional differential equations. This is because PINN heavily relies on automatic differentiation, and the classical chain rule applies to integral calculus and does not extend to fractional calculus. Nonetheless, a fractional form of the chain rule may be taken into consideration, but its application will be computationally costly. So, in the fPINN methodology, the derivatives that are integral order obtained from neural network approximation are derived using automatic differentiation, and the derivatives that are fractional order are numerically approximated using some conventional techniques that can discretize fractional operators numerically. The effectiveness of fPINN is validated for fractional advection-diffusion equations for both forward and inverse problems in one, two, and three dimensions. Zhang et al. \cite{rev78} utilized PINN to solve stochastic differential equations. For the same PINN, it was combined with the spectral dynamically orthogonal method and the bi-orthogonal method and coined NN-DO/BO. The DO/BO restrictions are added to the loss function in addition to the modal decomposition of the stochastic differential equation. The proposed method is said to have various advantages over the orthogonal method and the bi-orthogonal method, but it comes with a few limitations of PINN, including limited precision and computational time. However, if parallel implementation can be done, then the computational cost can be handled. Another work to solve stochastic differential equations was carried out by Yang et al. \cite{rev79}, where a physics-informed generative adversarial network was utilized. The method was demonstrated to be highly effective in solving high-dimensional problems.

Doleg{\l}o et al. \cite{rev30} explores the capability of NN in approximating the coefficient of the B-spline basis functions. In their work they present and compare three methods in solving one dimensional heat transfer problem. One, is the standard PINN where the solution is approximated using PINN and the residual is incorporated in the loss function. Secondly, the B-spline coefficients are approximated using neural network and the error function is roughly defined as the squared difference between neural network approximation and the coefficients obtained using Isogeometric analysis. Thirdly, the solution is approximated using neural network and the error function is roughly defined as the squared error between the neural network approximation and the one obtained using Isogeometric analysis. Second and third method is more like a supervised methodology of neural networks. It was concluded that neural network could approximate the coefficients accurately and efficiently in comparison to other two methods.

Guo et al. \cite{rev3} applied PINN to solve three PDEs and obtained good approximation results. In order to enhance the training process, they further employ a residual-based adaptive refinement technique, which involves adding more residual points in regions with high PDE residuals until the residuals reach below a certain threshold.

Baymani et al. \cite{rev7} utilized the neural network approximation to solve the Stokes problem. They approximated the solution using a trial function, in which one term satisfied the boundary conditions and the other term was the neural network approximation. The solution to the Stokes problem is obtained by converting it to three Poisson problems. Their examples clearly demonstrated that an accurate solution could be reached with fewer model parameters.

Blechschmidt and Ernst \cite{rev12} explored three different machine learning techniques to study high-dimensional PDEs and demonstrated the versatility of the methods. The three methods discussed are PINN, a technique related to the solution of backward stochastic DEs, and the technique related to the to the Feymann-Kac formula. In their work, they have also reviewed other machine learning techniques available in the literature. They have concluded that, among the three approaches, PINNs are the most appropriate for low-dimensional but non-linear PDEs. that are more complex in nature.

Waheed et al. \cite{rev35} utilized PINN to solve the eikonal equation. The flexibility of PINN to incorporate any additional physics makes the method stand out from the existing numerical techniques. To improve the method further, they have also explored the abilities of transfer learning. While using conventional techniques, minor changes in the model might require the same amount of effort for computation. The abilities of transfer learning come here as saviors. Additionally, for the precision of the solution and enhancement of the rate of convergence, an adaptive weighing strategy for the weights of the loss function and the adaptive activation function were incorporated.

Cheung and See \cite{rev41} used PINN to solve an advection diffusion equation and demonstrated a step-by-step solving process using the SimNet library by NVIDIA, which is a PDE solver based on PINN. The work also discusses and compares another PDE solver named the Fourier neural operator, which is a machine learning technique that makes use of solution information while training the neural network; unlike the PINN, it is a supervised methodology. One of the advantages of this method is that when the parameters are altered, the neural network does not require retraining, making it suitable for addressing parametric PDEs and those problems where the recovery of parameters is necessary.
\section{Application of PINNs in thermal management \& computational fluid dynamics}
PINNs have become a transformative approach for intricate fluid flow problems. Traditional computational fluid dynamics methods, while powerful, often require extensive computational resources and can be limited by the complexity of the governing equations and boundary conditions. On the other hand, PINNs use neural networks to explicitly embed physical principles into the learning process, enabling more accurate and efficient simulations of fluid flow phenomena. This section examines the PINN applications in the context of fluid dynamics, emphasizing recent advancements and methodologies where PINNs have been employed.

Rao et al. \cite{rev15} solved Newtonian laminar flows over a circular cylinder using mixed-variable PINN.  The proposed method converts the fundamental equations to continuum and constitutive forms, and considers the stream function to guarantee that the divergence-free condition is satisfied. The major advantage of the conversion is that the order of  derivatives is reduced which leads to the gain in the trainability and predictive accuracy of neural networks. Using a mixed variable method, the velocity and pressure field were accurately generated. It was shown that trainability of the model and the accuracy of the solution is improved when mixed variable scheme is employed. The same methodology was applied to solve laminar flow having heat transfer in a domain that is rectangular with two tube like obstructions \cite{rev16}. An analysis of the influence of the neural network architectures on the solution obtained using the proposed method is also carried out. In addition, normalization of the conservation equations is also done. This is done because, the magnitude of different terms differ largely which causes the loss function to be biased towards one single term leading to an increased training cost and wrong predictions.

Almajid et al. \cite{rev26} applied PINN to solve the Buckley-Leverett problem. Observed data was also incorporated into the loss function in the study. It was pointed out that the solution trend was obtained correctly using PINN without any observed data; however, by including the observed data, there was a noticeably large improvement in the solution accuracy. Observed data are generally data obtained from experiments or the available solution of the DEs. A detailed analysis was also conducted for different scenarios by including and excluding observed data and diffusion terms and by keeping multiphase parameters either trainable or non-trainable.

Jin et al. \cite{rev27} investigated the performance of PINN in solving incompressible flows. Navier Stokes flow nets (NSFnets) were formulated in two different forms: velocity-pressure form, in which the outputs are velocity and pressure, and velocity-voriticity form, in which the outputs are velocity and vorticity. In the former case, data for the boundary or initial condition for pressure is not available and is determined indirectly using the incompressibility constraint. On the other hand, in the latter case, the pressure term is removed and the incompressibility constraint is identically met. The methodology was also improved by utilizing a dynamic weighing strategy, transfer learning, and residual-based adaptive refinement.  When using PINNs, transfer learning enables us to initialize the network using the output of a previous training that solved the same PDE using a different source term. For example, a properly trained NSFnets with a lower Reynolds number can have its parameters transferred to the NSFnets with a higher Reynolds number.

Zhu et al. \cite{rev33} explored PINN to estimate the temperature, velocity, and pressure field for the metal additive manufacturing process. A soft-type approach and a hard-type approach were deployed to address two different sets of problems arising in metal manufacturing. The use of PINN in such situations is crucial, given the difficulty in generating data sets due to the high computing time required for simulations and the high cost of conducting experiments. The soft type approach is the conventional PINN approach, in which the boundary condition is enforced as an additional term in the loss function, whereas the Dirichlet boundary condition is exactly satisfied in the hard type approach by employing a heavyside function in the neural network approximation. When using a soft-type approach, there is no assurance that the neural network can approximate the boundary condition precisely. Also, the effectiveness of learning depends on the weight assigned to the boundary condition term. However, the hard-type approach is seen to enhance the learning process.

PINNs were also used to solve various heat transfer problems of industrial relevance by Cai et al. \cite{rev34}. They also took into account ill-posed problems with unknown thermal boundary conditions. In most of the applications, specifically in the field of power electronics, because of the need for intricate equipment, it is often very challenging to obtain the thermal boundary conditions. Therefore, for such scenarios, it is important to have a numerical technique that can handle missing conditions and provide reliable results. Also, using any CFD techniques to solve problems such as having measurements of temperature at a few points but having missing boundary conditions is a tedious process, but PINNs is a flexible method to use in such scenarios. In their work, they have also proposed a method that is capable of choosing locations for the placement of sensors adaptively. Their study emphasizes how we can make use of the availability of data that can be obtained using multifidelity methods and combine it with the governing equations with missing information so that effective computational techniques and experimental heat transfer can coexist.

Tartakovsky et al. \cite{rev32} implemented PINN to find hydraulic conductivity in both saturated and unsaturated flows. Applications with sparse data and partial information of physics were the main focus of the study. Due to the heterogeneous nature of porous media and the dearth of measurements for subsurface transport flow problems, it is extremely difficult to assimilate data for parameter estimation. Hence, methods capable of including physics and sparse measurements play a significant role in such applications. In their work, the hydraulic conductivity and hydraulic head are approximated using neural networks, and these are trained by incorporating Darcy’s equation as well as the hydraulic head and hydraulic conductivity measurements. He et al. \cite{rev36} extended this method to approximate the hydraulic conductivity, hydraulic head, and solute concentration using neural networks and trained these by incorporating sparse measurements of these with advection-dispersion equations and Darcy’s equation. The term multiphysics-informed neural network approach (MPINN) was used to describe this method. Their numerical results demonstrate that, when compared to the neural network method that purely relies on data, both of these PINN-based methods gave better accuracy; additionally, among the PINN approaches, MPINN gave superior parameter and state estimation.

Considering the lack of availability of data in parametric fluid dynamics problems, Sun et al. \cite{rev38} applied PINN without relying on any data set to solve a few fluid flow problems having applications in hemodynamics. The work deals with a problem that is challenging to solve using existing techniques, that is, problems defined over irregular geometries. The soft-type and hard-type approaches for imposing boundary conditions were also implemented and compared. It was concluded that enforcing the hard type boundary condition is crucial in learning data-free PINN to guarantee the learned solution is the unique solution of the PDE. The impact of adaptive activation function was also studied, and it was observed that for the accuracy of the solution, imposing hard-type boundary conditions is more significant than utilizing adaptive activation function. The study provided a promising result for the development of data-free surrogate fluid flow models.

Raghu et al. \cite{rev45} applied PINN to solve free shear flow, specifically steady jet flows. PINN is used to determine the mean flow variables; for the same, the equations considered are Reynolds-Averaged Navier Stokes equations. For more accurate predictions, a new architecture was developed without the need to rely on any data set. In the new architecture, two neural network models were taken into account. Mean velocity and pressure are the outputs from the first neural network, and the inputs to the next neural network are the velocity gradients, which output the viscosity. And to compute the Reynolds stresses, they used the Boussinesq hypothesis. The predictive ability of PINN is observed to enhance utilizing this architecture. In addition to that, owing to the poor convergence, an extended dynamic weighing strategy is also introduced. In the extended dynamic weighing strategy, in addition to using dynamic weights in the initial or boundary conditions, dynamic weights were introduced in individual terms of the loss, like the x-momentum equation, the z-momentum equation, etc. In this way, the gradients of loss for each PDE term are also distributed, resulting in better performance. It was also pointed out that when comparing the dimensional and non-dimensional forms of PDEs that have to be solved by PINN, the non-dimensionalized version performed better.

Dieva et al. \cite{rev46} solved fluid flows in porous media, which are non-Newtonian in nature, using PINN. The model was designed by minimizing the PINN loss function, which also contained the mean squared error of the data obtained from experiments for solving direct and inverse problems. Also, for the problem of study considered, an examination of the PINN's sensitivity to variations in the input parameter set reveals that the method is not sensitive to such changes.

Many different phenomena, including fluid flow, are frequently modelled using dynamic PDEs. Rout \cite{rev56} explored the application of PINN in solving dynamical PDEs. The importance of optimization algorithms and weights of loss functions is stressed in the paper for solving dynamical PDEs. To overcome the challenges of using PINN, major reasons for the bad performance of PINN when it comes to dynamical PDEs were discussed, which were determined with the help of case studies. The first is advection dominance, which is the phenomenon wherein the advection term in a differential equation has a greater influence than the diffusion term, posing difficulties for discrete approximation. The second is time variance. The use of a physics-informed recurrent neural network is suggested to be an ideal option when it comes to solving dynamical PDEs.

Bararnia and Esmaeilpour \cite{rev57} solved three benchmark problems arising in the boundary layer and thermal flow problems using PINN. The impact of the unbounded domain and the nonlinearity of the equation on the network architecture is studied in detail. It was also shown that the numerical value that is assumed for boundary conditions has to be bigger than the thickness of the boundary layer to get good results using PINN.

While performing experiments in fluid mechanics, precisely obtaining the value of velocity close to the wall is often challenging. So, with the availability of a few data points that are distant from the wall and utilizing PINN, flow predictions were obtained near the walls in the work of Sekar et al. \cite{rev58}. Additionally, the impact of sample point position on accuracy is examined. It was also observed that increasing Reynolds numbers make it difficult for PINN to reliably acquire solution close to the wall. 

In the study conducted by Aygun and Karakus \cite{rev59}, the use of PINNs in various thermal convection regimes to solve coupled fluid flow and heat transfer problems was done. The importance of the weights in terms of loss function and their influence on the solution accuracy is studied. The authors pronounced the importance of these weights for the different test cases that were considered in the study. They also considered the problems relevant to the power electronics industry. In their work, a few observation data points were used for improved accuracy. Additionally, rather than using a fully connected neural network to approximate the solution, different models that are present in the framework, NVIDIA Modulus, were also used, and their performances were compared. Other considered models include the Fourier network, the modified Fourier network, the deep Galerkin method, and the modified highway network. It was concluded that the test cases considered did not have multi-scale behaviours, so there is no requirement for Fourier mapping. Therefore, fully connected networks were experiencing faster convergence, and the Fourier mapping-based architectures converged slowly.

Chandra and Das \cite{rev63} solved coupled ODEs arising from the bioconvective boundary layer flow of nanofluids over a stretching surface. They proposed PINN combined with the stochastic algorithm, namely the interior point algorithm, which was used to optimize the weights. Statistical analysis is also carried out to show the reliability of the proposed method. The study has also examined the effects of two major parameters: the thermophoresis parameter and the Brownian motion parameter. Their method is observed to perform better than the traditional methods, but it lacks efficiency in optimization. It is noted that further improvement in the optimization strategy, like introducing hybrid methods, might lead to improved performance.

Butt et al. \cite{rev64} studied Williamson nanofluid flow over a stretching sheet. The solution was approximated using neural networks, and the inverse multiquadric radial basis function was utilized as an activation function. The optimization function utilized was the combination of genetic algorithms and sequential quadratic solver. By utilizing these, a novel design was proposed to analyse the magnetohydrodynamic flow of the viscoelastic fluid. It is stressed that the chosen activation, having the properties of probabilistic behaviour, radial symmetry, smooth transformation, and interpolation ability, comes with certain advantages, including its flexibility to various data distributions, easy adaptability to complex patterns, rapid learning of nonlinearity, and convergent learning process, respectively. Additionally, genetic algorithm is claimed to be the most reliable global search algorithms for optimization, and sequential quadratic programming is claimed to be one of the most reliable local search algorithm for achieving quick convergence for stiff nonlinear problems.

Chen et al. \cite{rev66} use WaveNets for the full-field recovery of rotating flow under nonlinear periodic water waves. WaveNets is the name formulated for solving the above problem using PINN. It consists of two neural networks; one of the networks outputs the wave elevation function, and the other outputs the velocity and pressure fields. In the method, a few data measurements are also used. In most studies that have incorporated PINNs, the governing equations are defined on a fixed domain. So, we randomly select the points for computing the residuals. However, in the case of the considered problem, only those points under the water surface need to be considered. As a result, in order to address this, which frequently occurs for free-surface problems, a novel approach to dynamically finding the residual points is presented. This method of dynamically choosing residual points can be applied to any problem with complex geometries. WaveNets gives good accuracy by making use of less available data for one-layer rotational flows as well as for two-layer rotational flows.

PINN has been applied to solve the transient unconfined groundwater flow \cite{rev80}. Both homogeneous isotropic and heterogeneous anisotropic aquifer instances were taken into consideration. This study is important since the governing PDE of the problem has a boundary condition that is spacetime-varying, which makes the problem challenging to handle using other techniques. From the obtained results, it can be concluded that PINN provides precise results and is also an ideal methodology for applications in hydrology. PINNs are especially helpful in scenarios where data availability is constrained. However, in the study, the accuracy and optimal choice of hyperparameters still remain a challenge. 

The nature of phase interfaces was studied by Jalili et al. \cite{rev81} in solving two-phase flows. A data-dependent PINN model was used for the same. The study examines how a single gas bubble rises within a dense fluid and how the heat transfers towards the heated wall. This work is significant as it shows the success of applying PINN to forward, inverse, and extrapolative multiphase problems. PIIN’s extension to analyse phase change that is temperature-dependent is a study to explore.

A brief discussion of a few of the variations of data-driven PINN model that are applied to fluid flow problems is provided below. Data-driven PINNs incorporates empirical data directly into the training process, along with the physics-based constraints. Unlike classical PINNs that rely solely on the governing equations, data-driven PINNs simultaneously optimize for both the physical laws and fit to observed data. Integrating these two information sources allows data-driven PINNs to model intricate processes, such fluid flow, for which empirical data can offer vital insights.

Thakur et al. \cite{rev74} introduce ViscoelasticNet, a PINN framework to study the distribution of stress. This framework uses the velocity field to select the constitutive relationship of the non-Newtonian fluid model and to study the stress field. The distribution of stress and velocity was further used to learn the distribution of pressure. The study utilized sparse data, and the results indicated that the approach could effectively determine the viscoelastic model parameters. Mahmoudabadbozchelou et al. \cite{rev73} present an approach based on PINN that also utilizes additional data to study the complex fluid independent of the kind of rheological constitutive model. This was coined the name Non-Newtonian Physics-Informed Neural Networks. Numerical experiments, including many models and flow regimes, were conducted, encompassing the application of problems with missing boundary conditions. The results of the study clearly imply that flow simulations of complex fluids can significantly benefit from the use of the proposed methodology. Another research effort advancing along similar lines is the introduction of rheology-informed neural networks \cite{rev75}. The proposed data-driven methodology is used to solve direct and inverse problems of rheological constitutive models that are often complicated. The methodology was applied to thixotropic, elastic, viscose, and plastic fluid flow in several flow regimes. However, it is not restricted to any specific fluid models and can be used to study different models. It was demonstrated that regardless of the kind of experimental data available, it is possible to easily and accurately calculate the model parameters. Hu et al. \cite{rev76} introduce a new methodology combining PINN with the characteristic-based split method, which is said to enhance the speed of PINN. This is because, using this methodology, all of the partial derivatives do not take part in gradient backpropagation. This method is applied to solve shallow water equations and Navier-Stokes equations. It should be noted that the methodology can handle data-free problems. With the help of numerical examples, it was shown that, even with a small data set, the method yields accurate results.

The study of fluid flow across a curved surface is the focus of the study conducted by Ganga et al. \cite{rev83}, which has applications in the polymer industry, which is crucial to the production of contact lenses. The work investigates the applicability of the method proposed in \cite{rev44} to solve fluid flow problems. It is investigated if the wavelet activation function can be used in PINN to solve nonlinear coupled differential equations in fluid flow problems. In addition, instead of using a single neural network model to approximate all functions, the study employed separate neural network models for different solutions simultaneously to estimate unknown functions. The governing differential equations are solved without relying on any additional simulation or experimental dataset. The non-Newtonian Maxwell fluid flow across a curved surface is formulated mathematically, taking into account the effects of several factors. It is observed from the analysis of various flow parameters that the proposed methodology gives reliable results. This work is important because it provides a viable path for implementing the proposed approach, which provides a simple way to address complex fluid flow problems. Nevertheless, a significant challenge remains in hyperparameter tuning.
\section{Future Directions}
PINNs have revolutionized the way we approach complex physical problems by integrating physical laws into neural network frameworks. This article highlights the evolution of PINN, the potential of PINN modifications to address certain limitations, and its applications to fluid flow problems. This may open up new avenues for PINN research in fluid dynamics and other fields. Despite the significant advancements made, there are still plenty of opportunities for further investigation. Some of the observations are listed below:\\
Application of PINN real-world problems: Simulating the physical systems of real-world phenomena presents a number of difficulties, such as high-dimensional and geometrically complex problems. There is a lack of PINN studies on industrial applications, and further investigation on how well PINNs can handle complex multi-physics and multi-scale applications can be carried out.\\
Theoretical analysis: For PINNs, theoretical analysis such as convergence and generalization capabilities are still in their early stages. Because it is hard to analyze the neural network training process, it remains a difficult task. Moreover, further mathematical foundations in optimization and numerical analysis will be needed for the new PINN techniques. This analysis may drive the development of more effective methods and architectures.\\
Implementation: More research is necessary to achieve an effective automated tuning of PINN implementation, as the selection of an appropriate architecture is problem-specific. Furthermore, it has been shown that establishing boundary conditions explicitly works better; nonetheless, the application of hard-type boundary conditions to any defined geometry can be further explored. Convolutional and recurrent neural network designs may also be further investigated in terms of their implementation and theoretical understanding.\\
Optimization: PINN's optimization strategies still have opportunity for development, given the tremendous effort that has been put into tackling optimization problems. The development of new and efficient optimisation methods is a promising avenue for future study.\\
Training time: A primary drawback of PINN is the need to train the neural network, which might need a notably longer duration than other widely used numerical techniques. In order to reduce training times, research into creating pre-trained models or transfer learning strategies might be explored.\\
Automatic differentiation: When computing higher-order derivatives, automatic differentiation might be prone to being ineffective and numerically unstable. However, not much research has been done on alternatives to automatic differentiation, which can be further investigated.
\bibliographystyle{elsarticle-num}
\bibliography{review_ref}
\end{document}